\documentclass[prd,twocolumn,showpacs,amsmath,amssymb]{revtex4}
\usepackage{graphicx}
\usepackage{dcolumn}
\usepackage{bm}
\usepackage{psfig}
\usepackage{epsfig}
\usepackage{overpic}
\usepackage{verbatim}
\usepackage[colorlinks,linkcolor=blue,anchorcolor=red,citecolor=blue]{hyperref}

\usepackage{rotating}

\newcommand{\bfg}{\begin{figure}}
\newcommand{\efg}{\end{figure}}
\newcommand{\bitm}{\begin{itemize}}
\newcommand{\eitm}{\end{itemize}}
\newcommand{\bnum}{\begin{enumerate}}
\newcommand{\enum}{\end{enumerate}}
\newcommand{\btbl}{\begin{table}}
\newcommand{\etbl}{\end{table}}
\newcommand{\btbu}{\begin{tabular}}
\newcommand{\etbu}{\end{tabular}}

\newcommand{\beq}{\begin{equation}}
\newcommand{\edq}{\end{equation}}

\begin{document}

\normalsize
\parskip=5pt plus 1pt minus 1pt

\title{\boldmath Cross section measurements of $e^{+}e^{-} \to  K^{+}K^{-}K^{+}K^{-} $ and $\phi K^{+}K^{-}$ at center-of-mass energies from 2.10 to 3.08~GeV}

\author{M.~Ablikim$^{1}$, M.~N.~Achasov$^{10,d}$, P.~Adlarson$^{59}$, S. ~Ahmed$^{15}$, M.~Albrecht$^{4}$, M.~Alekseev$^{58A,58C}$, A.~Amoroso$^{58A,58C}$, F.~F.~An$^{1}$, Q.~An$^{55,43}$, Y.~Bai$^{42}$
% ...
\\
(BESIII Collaboration)\\
$^{1}$ Institute of High Energy Physics, Beijing 100049, People's Republic of China\\
% ...
$^{4}$ Bochum Ruhr-University, D-44780 Bochum, Germany\\
% ...
$^{10}$ G.I. Budker Institute of Nuclear Physics SB RAS (BINP), Novosibirsk 630090, Russia\\
% ...
$^{15}$ Helmholtz Institute Mainz, Johann-Joachim-Becher-Weg 45, D-55099 Mainz, Germany\\
% ...
$^{42}$ Southeast University, Nanjing 211100, People's Republic of China\\
$^{43}$ State Key Laboratory of Particle Detection and Electronics, Beijing 100049, Hefei 230026, People's Republic of China\\
% ...
$^{55}$ University of Science and Technology of China, Hefei 230026, People's Republic of China\\
% ...
$^{58}$ (A)University of Turin, I-10125, Turin, Italy; (B)University of Eastern Piedmont, I-15121, Alessandria, Italy; (C)INFN, I-10125, Turin, Italy\\
$^{59}$ Uppsala University, Box 516, SE-75120 Uppsala, Sweden\\
% ...
$^{d}$ Also at the Novosibirsk State University, Novosibirsk, 630090, Russia\\
% ...
}

\date{\today}

\begin{abstract}

  We measure the Born cross sections of the process $e^{+}e^{-} \to
  K^{+}K^{-}K^{+}K^{-}$ at center-of-mass (c.m.) energies,
  $\sqrt{s}$, between 2.100 and 3.080~GeV. The data were collected
  using the BESIII detector at the BEPCII collider.
  An enhancement at $\sqrt{s}= 2.232$~GeV is observed, very close to
  the $e^{+}e^{-} \to \Lambda \overline{\Lambda}$ production threshold.
  A similar enhancement at the same c.m.\ energy is observed in the
  $e^{+}e^{-} \to \phi K^{+}K^{-}$ cross section.
  The energy dependence of the $K^{+}K^{-}K^{+}K^{-}$ and $\phi K^{+}K^{-}$ cross sections
  differs significantly from that of $e^{+}e^{-} \to \phi \pi^{+}\pi^{-}$.
\end{abstract}


\maketitle

\section{Introduction}

The $\phi(2170)$ resonance, denoted previously as $Y(2175)$, was first
observed by BABAR in the process $e^{+}e^{-} \to \phi f_{0}(980) \to
\phi \pi \pi$~\cite{Y2175BABAR11} via initial-state radiation (ISR) and was
confirmed by Belle~\cite{Y2175BELLE}. BES~\cite{Y2175BESII} and
BESIII~\cite{Y2175BESIII, Y2175BESIII2019} also observed the $\phi(2170)$
in the $\phi f_{0}(980)$ invariant-mass spectrum. The discovery of $s\bar{s}$ bound
states is of interest for the understanding of the strangeonium
spectrum, which is less well understood than for
example the hidden-charm states ($c\bar{c}$). The CLEO Collaboration
found the first evidence for $Y(4260) \to
K^{+}K^{-}J/\psi$~\cite{Y4260} above the $D\bar{D}$-production
threshold. A similar process, $e^{+}e^{-} \to \phi K^{+}K^{-}$,
potentially allows the study of strangeonium-like vector states above the
$K\bar{K}$-production threshold.

Many theoretical interpretations have been proposed for the
$\phi(2170)$, such as a $s \overline{s}g$ hybrid~\cite{Y2175hybrid}, a
$2^{3}D_{1}$ $s\overline{s}$ state~\cite{Y2175ss2D}, a tetraquark
state~\cite{Y2175tetraquark1,Y2175tetraquark2}, a
$\Lambda\overline{\Lambda}$ bound state~\cite{Y2175lambda, Y2175lambda2}, or a
three-meson system $\phi K^{+}K^{-}$~\cite{X2170}.  The $1^{--}$
$s\overline{s}g$ hybrid can decay to $\phi \pi\pi$, with a cascade
$(s\overline{s}g \to (s\overline{s})(gg) \to
\phi\pi\pi)$~\cite{Others}, whereby $s\overline{s}g \to \phi
f_{0}(980)$ may make a significant contribution. However, none of the
theoretical models has so far been able to describe all experimental
observations in all aspects. Searching for new decay modes and
measuring the line shapes of their production cross sections will be
very helpful for interpreting the internal structure of the
$\phi(2170)$ resonance.

The BABAR Collaboration measured the $e^{+}e^{-} \to
K^{+}K^{-}K^{+}K^{-}$ cross sections and observed an enhancement
around 2.3 GeV~\cite{Y2175BABAR1, Y2175BABAR2}. In addition, the BES
Collaboration observed the $f_{0}(980)$, $f_{2}^{\prime}(1525)$ and
$f_{0}(1790)$ in the invariant-mass distribution of $K^{+}K^{-}$ pairs
in events in which the other $K^{+}K^{-}$ pair has an invariant mass
close to the nominal $\phi$ mass~\cite{JpsiToPhiKK}. An enhancement at
$\sqrt{s}= 2.175$~GeV was seen in the line shape of the process $e^{+}e^{-}
\to \phi f_{0}(980)$~\cite{Y2175BABAR2}, but due to poor
statistics, no strong conclusion could be drawn from the
data. Torres \emph{et al.} have performed a Faddeev calculation for the
three-meson system $\phi K^{+}K^{-}$ and obtained a peak around
2.150~GeV/$\it{c}^{\mathrm{2}}$~\cite{X2170}. These observations
stimulate 
experimentalists to study the energy dependence for the production of
the $\phi K^{+}K^{-}$ and $K^{+}K^{-}K^{+}K^{-}$ final states.

Using a data sample corresponding to an integrated luminosity of
650~$\mathrm{pb}^{-1}$ collected at center-of-mass (c.m.)~energies
from 2.0~GeV to 3.08~GeV~\cite{LuminosityFinal}, we present in this
paper the results of a study of the reaction $e^{+}e^{-} \to
K^{+}K^{-}K^{+}K^{-}$ and its dominant intermediate process
$e^{+}e^{-} \to \phi K^{+}K^{-}$.

\section{Detector and data samples}

The BESIII detector is a magnetic spectrometer~\cite{BESIII} located at the Beijing Electron Positron Collider (BEPCII)~\cite{BEPCII}.
The cylindrical core of the BESIII detector consists of a helium-based multilayer drift chamber (MDC), a plastic scintillator time-of-flight system (TOF),
and a CsI(Tl) electromagnetic calorimeter (EMC), which are all enclosed in a superconducting solenoidal magnet providing a 1.0~T magnetic field.
The solenoid is supported by an octagonal flux-return yoke with resistive plate counter muon identifier modules interleaved with steel.
The acceptance of charged particles and photons is 93\% over $4\pi$ solid angle. The charged-particle momentum resolution at $1~{\rm GeV}/c$ is $0.5\%$,
and the $dE/dx$ resolution is $6\%$ for the electrons from Bhabha scattering. The EMC measures photon energies with a resolution of $2.5\%$ ($5\%$) at $1$~GeV
in the barrel (end cap) region. The time resolution of the TOF barrel part is 68~ps, while that of the end cap part is 110~ps.

The optimization of event-selection criteria, the determination of
detection efficiencies and the estimates of potential backgrounds are
performed based on Monte Carlo (MC) simulations taking the various
aspects of the experimental setup into account. The {\sc
  geant4}-based~\cite{Geant4} MC simulation software, which includes
the geometric and material description of the BESIII detector, the
detector response and digitization models, and the detector running
conditions and performances, is used to generate the MC samples.

For the background study, the $e^{+}e^{-} \to q \bar{q}$ process is
simulated by the MC event generator {\sc conexc}~\cite{ConExc}, while
the decays are generated by {\sc evtgen}~\cite{EVENTGEN1,
  BESEVENTGEN2} for known decay modes with branching fractions set to
Particle Data Group (PDG) world-average values~\cite{PDG} and by {\sc
  luarlw}~\cite{LUARLW} for the remaining unknown decays. MC samples
of $e^{+}e^{-} \to e^{+}e^{-}$ and $\mu^{+}\mu^{-}$ processes are
generated by {\sc babayaga} 3.5~\cite{Babayaga}. The signal MC samples
from the phase-space models (PHSP) of $e^{+}e^{-} \to
K^{+}K^{-}K^{+}K^{-}$ and $e^{+}e^{-} \to \phi K^{+}K^{-}$ are
generated at c.m.\ energies corresponding to the experimental values,
where the line shape of the production cross section of the two
processes is taken from the BABAR experiment~\cite{Y2175BABAR2} and
the signal detection efficiency is obtained by weighting the
MC-generated PHSP sample to data according to the observed invariant-mass distribution.

\section{Event Selection and background analysis}

\subsection{\boldmath $e^{+}e^{-} \to K^{+}K^{-}K^{+}K^{-}$}

Candidate events are required to have three or four charged
tracks. Charged tracks are reconstructed from hits in the MDC within
the polar angle range $|\mathrm{cos}\mathrm{\theta}|<0.93$ and
are required to pass the interaction point within 10~cm along the beam
direction and within 1~cm in the plane perpendicular to the beam. For
each charged track, the TOF and the $dE/dx$ information are combined
to form particle identification (PID) confidence levels (C.L.)  for
the $\pi$, $K$, and $p$ hypotheses. The particle type with the highest
C.L. is assigned to each track. At least three kaons are required to
be identified. The primary vertex of the event is reconstructed by
three kaons. For events with four identified kaons, the combination
with the smallest chi-square of the vertex fit is
retained.

\begin{figure}[htbp]
\begin{center}
\begin{overpic}[width=7.5cm,height=5.5cm,angle=0]{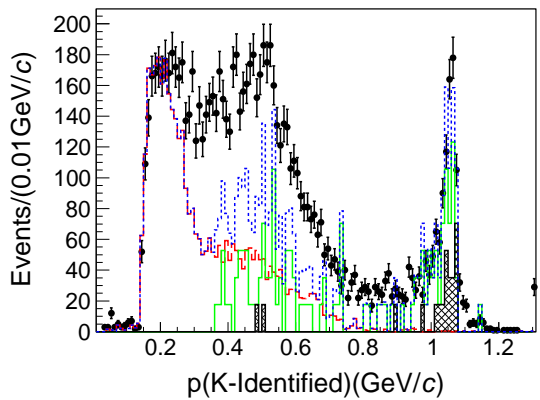}
\end{overpic}
\end{center}
\vspace*{-0.6cm}
\caption{(color online). Momentum spectrum of the three identified
  kaons at $\sqrt{s} = 2.125$~GeV. The black dots with error bars are
  data, the dashed (red) histogram is from $e^{+}e^{-} \to q\bar{q}$, the solid (green) histogram is from $e^{+}e^{-} \to e^{+}e^{-}$, the hatched (black)
  histogram is from $e^{+}e^{-} \to \mu^{+}\mu^{-}$, and the dotted (blue)
  histogram is the sum of all MC samples.  }
\label{MissingKaonFoundMomentumAllMC}
\end{figure}

Figure~\ref{MissingKaonFoundMomentumAllMC} shows the momentum
distribution of the three identified kaons for $\sqrt{s}$ = 2.125~GeV
after applying the above-mentioned selection criteria. The peak on the
right-side of the spectrum stems from reducible QED background,
dominated by the processes $e^{+}e^{-} \to e^{+}e^{-}$ and $e^{+}e^{-} \to
\mu^{+}\mu^{-}$. To suppress this background, the momenta of
the identified particles are required to be less than 80\% of the mean
momentum of the colliding beams ($p_\text{beam}$).
 
\subsection{\boldmath $e^{+}e^{-} \to \phi K^{+}K^{-}$}

For $e^{+}e^{-} \to \phi K^{+}K^{-}$ with $\phi \to K^{+}K^{-}$, the
final state is $K^{+}K^{-}K^{+}K^{-}$. The selection criteria for
three or four kaons are the same as described in the previous
subsection. In addition to the primary-vertex fit of the three kaons,
a one-constraint (1C) kinematic fit is performed under the hypothesis
that the $KK^{+}K^{-}$ missing mass corresponds to the kaon mass. For
events with four reconstructed and identified kaons, the combination
with the smallest chi-square of the 1C kinematic fit
($\chi^{2}_{1\mathrm{C}}(K^{+}K^{-}KK_\text{miss})$) is retained and
required to be less than 20.  In the following, the $K_\text{miss}$
momentum is that obtained from the 1C kinematic fit and is used in
invariant-mass calculations.

\begin{figure}[htbp]
\begin{center}
\begin{overpic}[width=7.5cm,height=5.5cm,angle=0]{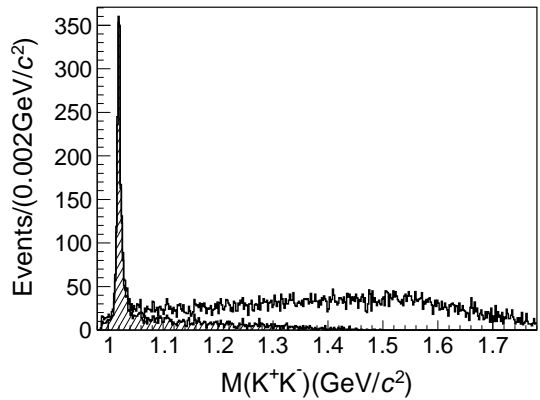}
\end{overpic}
\end{center}
\vspace*{-0.6cm}
\caption{(color online). Invariant-mass distribution at $\sqrt{s}=3.080$~GeV for all $K^{+}K^{-}$ pairs in selected $e^{+}e^{-} \to K^{+}K^{-}K^{+}K^{-}$ events (open histogram), and for the combination in each event closest to the $\phi$-meson mass (hatched). }
\label{PhiKKFourEntry3080}
\end{figure}

The open histogram in Fig.~\ref{PhiKKFourEntry3080} shows the invariant-mass distribution for all $K^{+}K^{-}$ pairs for the
selected $K^{+}K^{-}K^{+}K^{-}$ events (four entries per event) for data taken at $\sqrt{s} =  3.080$~GeV.
The hatched histogram in the same figure corresponds to the distribution of the pair with a mass closest to the nominal $\phi$ mass.
A prominent peak near the $\phi$ mass is seen in both histograms and indicates that the $\phi~K^{+}K^{-}$ channel dominates the $K^{+}K^{-}K^{+}K^{-}$ final states.

\section{Signal yields}

\begin{figure}[htbp]
\begin{center}
\begin{overpic}[width=9.0cm,height=6.5cm,angle=0]{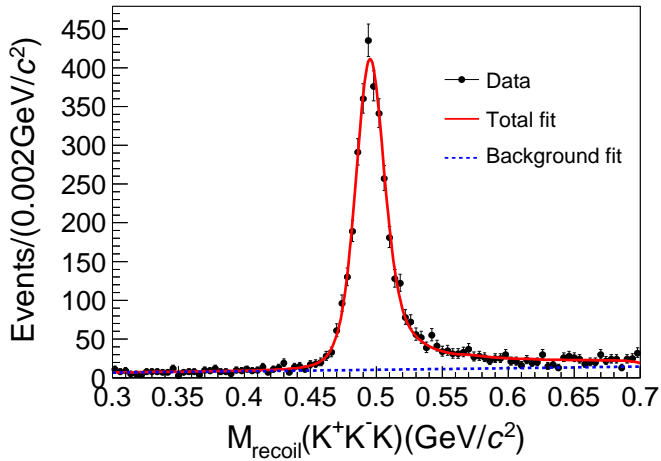}
\end{overpic}
\end{center}
\vspace*{-0.6cm}
\caption{(color online). The fit to the
  $M_{\mathrm{recoil}}(K^{+}K^{-}K)$ mass spectra at $\sqrt{s}$ =
  3.080~GeV. The black dots with error bars are data, the solid (red)
  curve shows the result of the best fit, and the dashed (blue) curve
  shows the result for the background.}
\label{Fit3080KKKK}
\end{figure}

The signal yields of $e^{+}e^{-} \to K^{+}K^{-}K^{+}K^{-}$ are obtained from unbinned maximum-likelihood fits to the
$K^{+}K^{-}K$ recoil-mass ($M_{\mathrm{recoil}}(K^{+}K^{-}K)$) data. The signal is described by the line shape obtained from the MC simulation convolved with
a Gaussian function, where the Gaussian function describes the difference in resolution between data and MC simulation. The background shape is parametrized
by a second-order Chebyshev polynomial function. The parameters of the Gaussian function and the Chebyshev polynomial function are left free in the fit.
The corresponding fit result for data taken at $\sqrt{s}=3.080$~GeV is shown in Fig.~\ref{Fit3080KKKK}.

To determine the signal yields of the $e^{+}e^{-} \to \phi K^{+}K^{-}$ process, an unbinned maximum-likelihood fit is performed to the $M(K^{+}K^{-})$ spectra.
The probability density function of the $M(K^{+}K^{-})$ spectra for the $\phi$ is obtained from a $P$-wave Breit-Wigner function convolved with a
Gaussian function that accounts for the detector resolution. The $P$-wave Breit-Wigner function is defined as

\begin{equation}
\mathit {f}(m)=|\mathrm{A}(m)|^{2}\cdot p,
\end{equation}

\begin{equation}
\mathrm{A}(m)=\frac{p^{\ell}}{m^2-m^{2}_{0}+im\Gamma(m) }
\cdot\frac{\mathrm{B}(p)}{\mathrm{B}(p^{'})},
\end{equation}

\begin{equation}
\mathrm{B}(p)=\frac{1}{\sqrt{1+(\mathrm{R}p)^{2}}},
\end{equation}

\begin{equation}
\Gamma(m)=\left(\frac{p}{p^{'}}\right)^{2\ell + 1}\left(\frac{m_{0}}{m}\right)\Gamma_{0}\left[\frac{\mathrm{B}(p)}{\mathrm{B}(p^{'})}\right],
\end{equation}

\noindent where $m_{0}$ is the nominal $\phi$ mass as specified by the
PDG, $p$ is the momentum of the kaon in the rest frame of the $K^+K^-$
system, $p^{'}$ is the momentum of the kaon at the nominal mass of the
$\phi$, and $\Gamma_{0}$ is the width of the $\phi$. The angular
momentum ($\ell$) is assumed to equal one, which is the lowest allowed
given the parent and daughter spins, $\mathrm{B}(p)$ is the
Blatt-Weisskopf form factor, and $R$ is the radius of the centrifugal
barrier, whose value is taken to be 3~{GeV/\it{c}}$^{-1}$~\cite{Blatt-Wdisskopf}.

\begin{figure}[htbp]
\begin{center}
\begin{overpic}[width=9.0cm,height=6.5cm,angle=0]{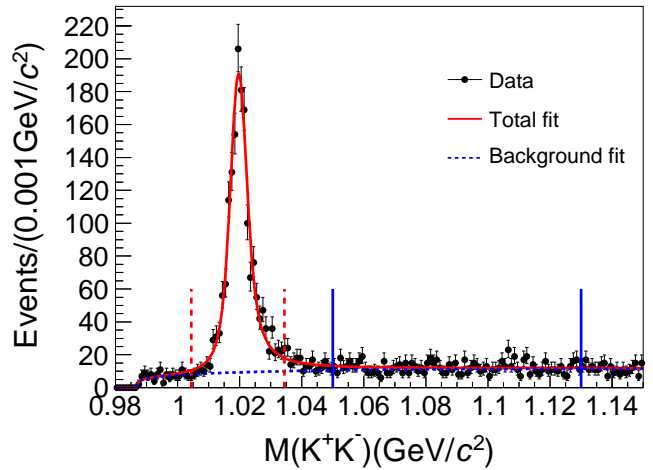}
\end{overpic}
\end{center}
\vspace*{-0.6cm}
\caption{(color online). Fit to the $M(K^{+}K^{-})$ mass spectrum
  (four entries per event) at $\sqrt{s} = 3.080$~GeV. The black dots
  with error bars are for data, the solid
  (red) curve represents the total fit result and the dashed (blue)
  curve corresponds to the background contribution determined by the
  fit. Also shown are the signal (vertical dashed red lines) and
  side-band (vertical solid blue lines) regions used for the determination of the $K^+K^-$ (non-$\phi$ pair)
  invariant-mass distributions in Fig.~\ref{FitPhiNormalization}.}%
\label{Fitting3080}
\end{figure}

The background shape is described by an ARGUS
function~\cite{ArgusFunction}. The parameters of the Gaussian function
and the ARGUS function are left free in the fit.  The corresponding
fit result for data taken at $\sqrt{s} = 3.080$~GeV is shown in
Fig.~\ref{Fitting3080}.

The same event selection criteria and fit procedure are applied to
the other 19 data samples taken at different c.m.~energies.  The
number of events for these samples are listed in
Tables~\ref{XSectionKpKmKpKm} and
\ref{XSectionSummaryPhiKK}. 


\section{Selection efficiency}

\subsection{ \boldmath $e^{+}e^{-} \to \phi K^{+}K^{-}$}

\begin{figure}[htbp]
\begin{center}
\begin{overpic}[width=7.5cm,height=5.5cm,angle=0]{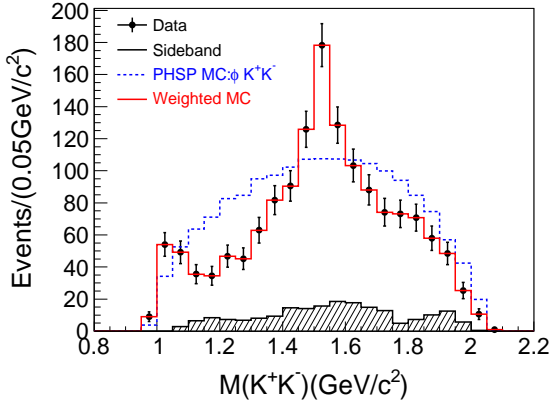}
\end{overpic}
\end{center}
\vspace*{-0.6cm}
\caption{(color online). Invariant-mass distribution of $K^{+}K^{-}$
  (non-$\phi$ pair) at $\sqrt{s}$ = 3.080 GeV. Here, the black dots
  with error bars are background-subtracted data, the hatched (black)
  histogram is the background determined from the $\phi$ side-band
  region, the dashed histogram is $\phi K^{+}K^{-}$ PHSP MC, and the
  solid (red) histogram is the weighted MC.  }
\label{FitPhiNormalization}
\end{figure}

The detection efficiency is obtained by MC simulations of the $\phi
K^{+}K^{-}$ channel using PHSP. It is found that data deviate strongly
from the PHSP MC distributions, as demonstrated by the histograms in
Fig.~\ref{FitPhiNormalization}, which show the non-$\phi$ pair
$K^+K^-$ invariant-mass distributions.  Here, $\phi$ candidates are
selected in the signal region and background from the side-band region
shown in Fig.~\ref{Fitting3080}.  The background in
Fig.~\ref{FitPhiNormalization} is the distribution of the invariant
mass of the remaining pair in the side-band event, and the data points
are the invariant mass of the remaining pair of the $\phi$ candidates
minus the background.  To obtain a more accurate detection efficiency,
the MC-generated events are weighted according to the observed
$K^{+}K^{-}$ (non-$\phi$ pair) invariant-mass distribution, where the
weight factor is the ratio of the $K^{+}K^{-}$ mass distribution between
data and PHSP MC. The weighted PHSP MC distribution is consistent with
the background-subtracted data, as shown by the solid histogram in
Fig.~\ref{FitPhiNormalization}.  The detection efficiencies determined
by using the weighted MC data and by using the $\phi K^{+}K^{-}$ PHSP
MC data do not differ significantly.  Therefore, the average detection
efficiency does not strongly depend on the $K^{+}K^{-}$ invariant
mass.

\subsection{\boldmath  $e^{+}e^{-} \to K^{+}K^{-}K^{+}K^{-}$}

The detection efficiency is determined using both the $\phi
K^{+}K^{-}$ weighted PHSP MC and $K^{+}K^{-}K^{+}K^{-}$ PHSP MC.  The
combined detection efficiency is given by

\begin{figure}[htbp]
\begin{center}
\begin{overpic}[width=7.5cm,height=5.5cm,angle=0]{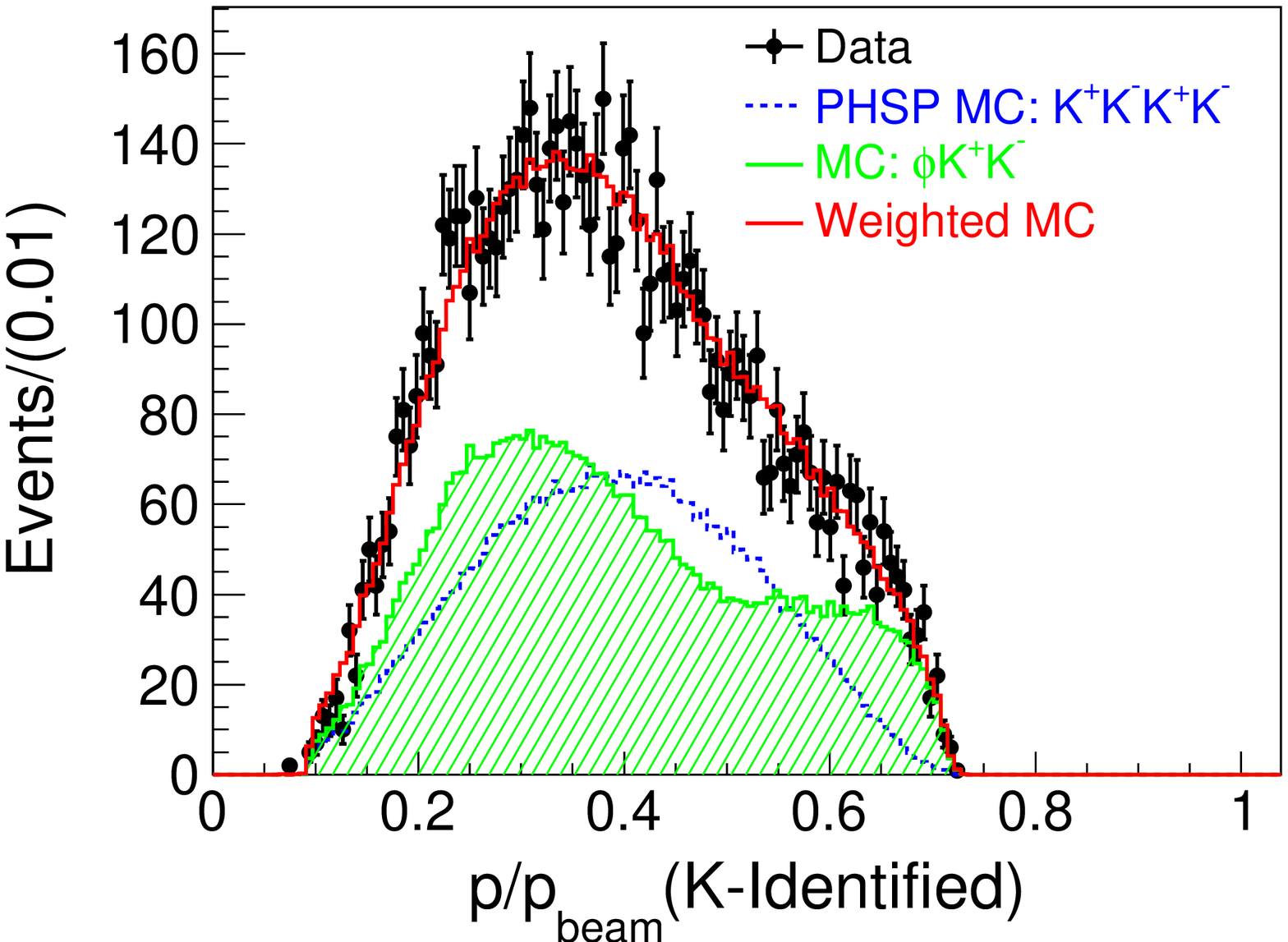}
\put(20,60){(a)}
\end{overpic}
\begin{overpic}[width=7.5cm,height=5.5cm,angle=0]{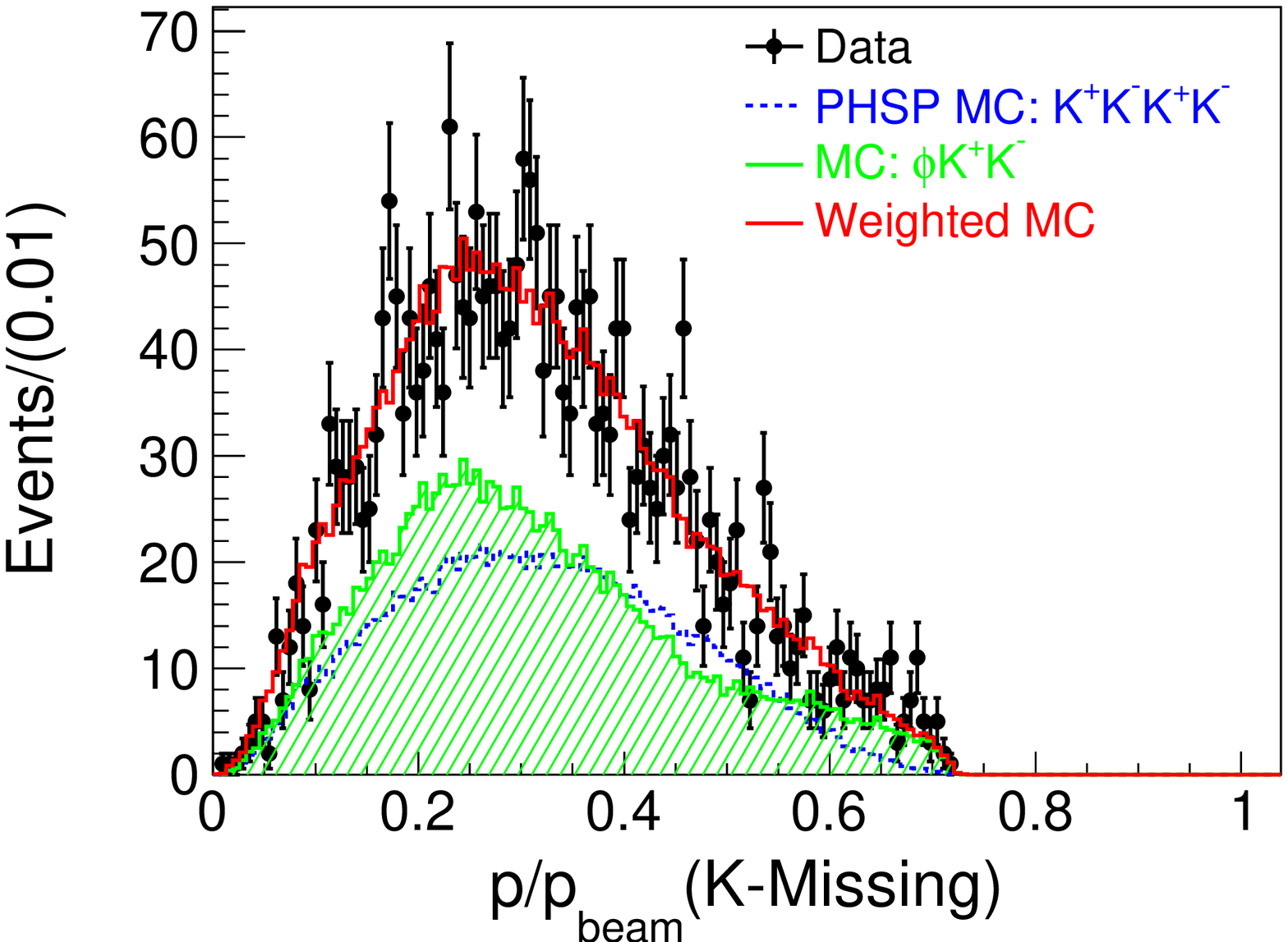}
\put(20,60){(b)}
\end{overpic}
\end{center}
\vspace*{-0.6cm}
\caption{(color online). (a) Normalized momentum spectra of three
  identified kaons (K-Identified) and (b) the recoiled kaon
  (K-Missing) of $e^+ e^- \to K^+ K^- K^+ K^-$ events at $\sqrt{s} =
  3.080$~GeV. Here, the black dots with error bars are data, the dashed
  (blue) histograms are $K^{+}K^{-}K^{+}K^{-}$ PHSP MC, the hatched
  (green) histograms are $\phi K^{+}K^{-}$ PHSP MC and the solid (red)
  histograms are the weighted MC samples. }
\label{Momen3080}
\end{figure}

\begin{center}
\begin{equation}
\mathrm{\epsilon}= \sum_{i=1}^{i=2}\omega_{i} \mathrm{\epsilon_{\mathit{i}}}  \quad {\rm with}   \quad  \omega_{i} = \mathit{N_{i}}/N_\text{total}.
\label{EfficiencyOfKKKK}
\end{equation}
\end{center}

\noindent where $\mathrm{\epsilon_{\mathit{i}}}$ and $N_{\mathit{i}}$
denote the detection efficiency and the signal yields of the
$\mathit{i}$th mode, respectively. $N_\text{total}$ is the total signal
yield obtained by fitting the $K^{+}K^{-}K$ recoil-mass data, $N_1$ is
the $\phi K^+K^-$ signal yield, $N_2 = N_\text{total} - N_1$, and
$\mathrm{\epsilon}$ is the weighted detection efficiency for the
$K^{+}K^{-}K^{+}K^{-}$ final state. Figure~\ref{Momen3080} shows a
comparison of the normalized momentum spectra of the kaon between the
data and the weighted MC result for $\sqrt{s}=3.080$ GeV.

\section{Determination of the Born cross section}

The Born cross section is calculated by
\begin{equation}
\mathrm{\sigma^{\mathit{B}}}=\frac{\mathit{N^{\text{obs}}}}{\mathcal{L} \cdot (1+\mathrm{\delta})
\cdot \mathrm{\epsilon}} ,
\label{EquationPhiKK}
\end{equation}
where $\mathit{N^\text{obs}}$ is the number of observed signal events,
$\mathcal{L}$ is the integrated luminosity, $(1+\mathrm{\delta})$
stands for $(1+\delta^{\mathrm{r}})\cdot(1+\delta^{\mathrm{v}})$, and
$(1+\delta^{\mathrm{r}})$ is the ISR correction
factor, which is obtained by a QED calculation~\cite{ISR} and by
taking the line shape of the Born cross section measured by the BABAR
experiment into account.  The vacuum polarization factor
$(1+\delta^{\mathrm{v}})$ is taken from a QED calculation with an
accuracy of 0.5$\%$~\cite{VP}, and $\mathrm{\epsilon}$ is the
detection efficiency. The branching fraction of the intermediate
process $\phi \to K^{+}K^{-}$ (49.2 $\pm$ 0.5$\%$)~\cite{PDG} is taken
into account in the determination of the cross section of $e^{+}e^{-}
\to \phi K^{+}K^{-}$.

Both $\mathrm{\epsilon}$ and $(1+\mathrm{\delta})$ are obtained from
MC simulations of the signal reaction for each c.m.\ energy. In the
{\sc conexc} generator, the cross section for the ISR process
($\sigma_{e^{+}e^{-}\rightarrow \gamma X}$) is parametrized using

\begin{equation}
\sigma_{e^{+}e^{-}\rightarrow \gamma X} = \int d\sqrt{s'}\frac{2\sqrt{s'}}{s}W(s,x)\frac{\sigma^{\mathit{B}}(\sqrt{s'})}{[1-\Pi(\sqrt{s'})]^2},
\end{equation}

\noindent
where $\sqrt{s'}$ is the effective c.m.\ energy of the final state with
$s' = s(1-x)$, $x$ depends on the energy of the radiated photon
according to $x = 2E_{\gamma}/\sqrt{s}$, $W(s, x)$ is the radiator
function and $\Pi(\sqrt{s'})$ describes the vacuum polarization (VP)
effect.  The latter includes contributions from leptons and
quarks. The detection efficiency and the radiative-correction factor
depend on the input cross section, and are determined by an iterative
procedure, in which the line shape of the cross section from BABAR is
used initially, and the updated Born cross section is obtained
according to the simulation. We repeat the procedure until the
measured Born cross section does not change by more than 0.5$\%$.

\begin{figure*}[htbp]
\begin{center}
\begin{overpic}[width=7.5cm,height=6cm,angle=0]{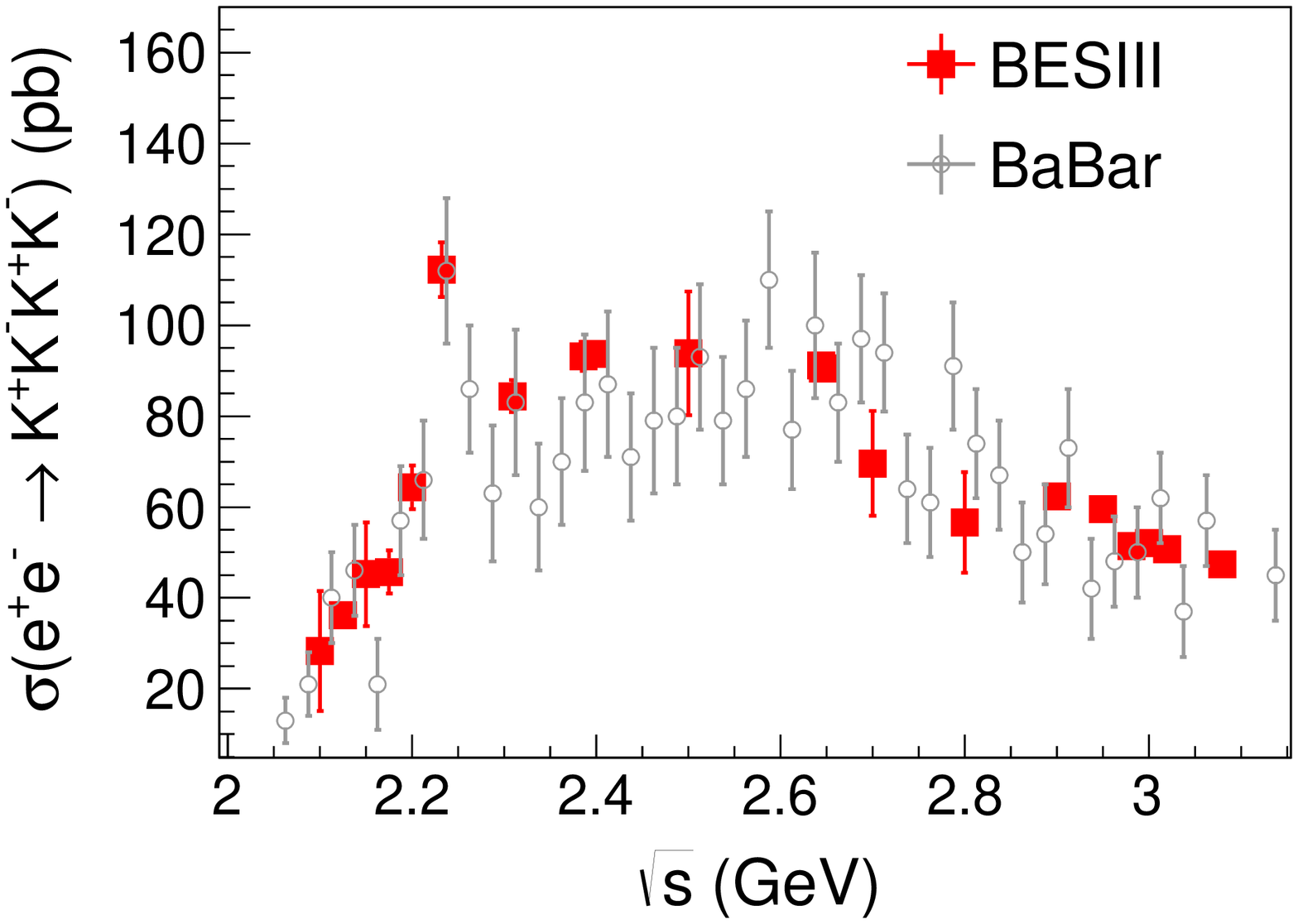}
\put(20,70){(a)}
\end{overpic}
 \begin{overpic}[width=7.5cm,height=6cm,angle=0]{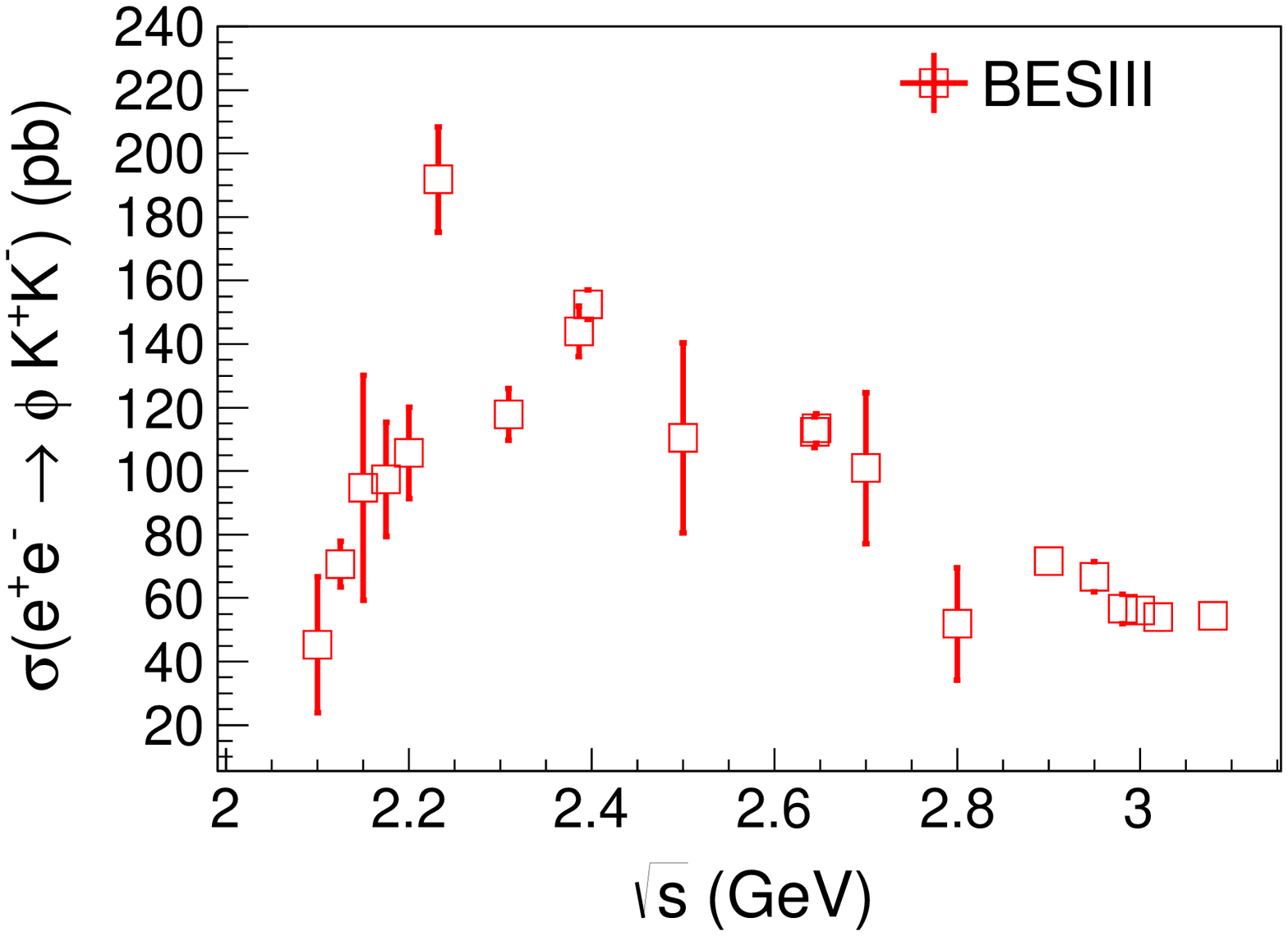}
\put(20,70){(b)}
\end{overpic}
\end{center}
\vspace*{-0.5cm}
\caption{(color online). (a) Comparison of the measured Born cross section of $e^{+}e^{-} \to K^{+}K^{-}K^{+}K^{-}$ to that of previous measurements~\cite{Y2175BABAR2}. The gray circles are from BABAR, the red rectangles are the results obtained in this work.  The BESIII results include statistical and systematical uncertainties. The errors of the BABAR data only include the statistical uncertainty. (b) Born cross section of $e^{+}e^{-} \to \phi K^{+}K^{-}$ obtained in this work. For BESIII data, the errors reflect both statistical and systematical uncertainties. }
\label{XSectionComparisonSummary}
\end{figure*}

The values of $\mathcal{L}$, $\mathit{N^\text{obs}}$, $(1+\delta)$
and $\mathrm{\epsilon}$ are listed in Table~\ref{XSectionKpKmKpKm},
together with the measured cross section at each energy point.
Figures~\ref{XSectionComparisonSummary} (a) and
\ref{XSectionComparisonSummary} (b) show the line shapes of cross
sections for $e^{+}e^{-} \to K^{+}K^{-}K^{+}K^{-}$ and $e^{+}e^{-} \to
\phi K^{+}K^{-}$, respectively.


\begin{table}[htbp]
  \centering
  \caption{The Born cross sections of $e^{+}e^{-} \to K^{+}K^{-}K^{+}K^{-}$. The center-of-mass energy ($\sqrt{s}$), integrated luminosity ($\mathcal{L}$), the yields of signal events ($\mathit{N^\text{obs}}$),  the product of radiative correction factor and vacuum polarization factor $(1+\mathrm{\delta})$, detection efficiency ($\mathrm{\epsilon}$), and Born cross section ($\sigma^{\mathit{B}}$). The first uncertainties are statistical and the second systematic.}
  
  \vskip5pt
  
 \newsavebox{\tablebox}
 \begin{lrbox}{\tablebox}

  \begin{tabular}{l r  r  r  r  r  r}
  \hline
  \hline
  $\sqrt{s}$ (GeV)  &$\mathcal{L}$ ($\mathrm{pb^{-1}}$)
    &$\mathit{N^\text{obs}}$ &  $(1+\mathrm{\delta})$  &$\mathrm{\epsilon}(\%)$  &$\mathrm{\sigma}^{\mathit{B}}$ (pb) &\\
 \hline
2.1000	  &12.2	  &18.9$\pm$8.8	  &0.8186	  &6.71	  &28.3$\pm$13.2$\pm$2.0\\
2.1250	  &109	  &378.7$\pm$19.3	  &0.8437	  &11.43	  &36.2$\pm$1.8$\pm$1.8\\
2.1500	  &2.84	  &18.3$\pm$4.6	  &0.8616	  &16.53	  &45.2$\pm$11.4$\pm$3.6\\
2.1750	  &10.6	  &95.6$\pm$9.9	  &0.8750	  &22.48	  &45.7$\pm$4.7$\pm$4.1\\
2.2000	  &13.7	  &206.6$\pm$15.3	  &0.8824	  &26.57	  &64.3$\pm$4.8$\pm$5.8\\
2.2324	  &11.9	  &369.2$\pm$19.8	  &0.8505	  &32.62	  &112.2$\pm$6.0$\pm$5.3\\
2.3094	  &21.1	  &682.3$\pm$28.0	  &0.9388	  &40.82	  &84.4$\pm$3.5$\pm$5.8\\
2.3864	  &22.5	  &934.6$\pm$32.0	  &0.9515	  &46.78	  &93.1$\pm$3.2$\pm$4.4\\
2.3960	  &66.9	  &2838.7$\pm$57.4	  &0.9534	  &47.53	  &93.7$\pm$1.9$\pm$7.2\\
2.5000	  &1.10	  &55.3$\pm$8.0	  &0.9741	  &55.13	  &93.8$\pm$13.6$\pm$5.3\\
2.6444	  &33.7	  &1819.9$\pm$47.0	  &1.0044	  &58.92	  &91.2$\pm$2.4$\pm$4.2\\
2.6464	  &34.0	  &1817.6$\pm$47.1	  &1.0049	  &58.77	  &90.5$\pm$2.3$\pm$4.1\\
2.7000	  &1.03	  &44.2$\pm$7.3	  &1.0173	  &60.40	  &69.6$\pm$11.5$\pm$6.2\\
2.8000	  &1.01	  &37.2$\pm$7.3	  &1.0424	  &62.50	  &56.6$\pm$11.1$\pm$3.7\\
2.9000	  &105	  &4366.4$\pm$76.1	  &1.0686	  &62.22	  &62.4$\pm$1.1$\pm$2.9\\
2.9500	  &15.9	  &629.1$\pm$29.5	  &1.0799	  &61.43	  &59.5$\pm$2.8$\pm$2.8\\
2.9810	  &16.1	  &555.6$\pm$28.1	  &1.0846	  &61.98	  &51.4$\pm$2.6$\pm$2.5\\
3.0000	  &15.9	  &557.3$\pm$28.1	  &1.0860	  &62.17	  &52.0$\pm$2.6$\pm$2.4\\
3.0200	  &17.3	  &591.4$\pm$29.2	  &1.0854	  &62.21	  &50.7$\pm$2.5$\pm$2.6\\
3.0800	  &126	  &3693.7$\pm$73.1	  &1.0185	  &60.59	  &47.4$\pm$0.9$\pm$2.2\\
 \hline
    \end{tabular}
    
    \end{lrbox}
    \scalebox{0.75}{\usebox{\tablebox}}
    
    \label{XSectionKpKmKpKm}
\end{table}

\begin{table}[htbp]
  \centering
  \caption{The same as Table~\ref{XSectionKpKmKpKm}, but for $e^{+}e^{-} \to \phi K^{+}K^{-}$.  Here, $\sigma^B$ is the cross section determined by Eq.~\ref{EquationPhiKK} divided by the branching fraction of $\phi \to K^+ K^-$.
  }

  \vskip5pt
 
 \begin{lrbox}{\tablebox}
 
  \begin{tabular}{l r  r  r  r  r  }
  \hline
  \hline
  $\sqrt{s}$ (GeV) &$\mathcal{L}$ (pb$^{-1}$)   &$\mathit{N^\text{obs}}$   &$(1+\delta)$  &$\mathrm{\epsilon}(\%)$   &$\sigma^{\mathit{B}}$ (pb)\\
  \hline
2.1000	 &12.2	 &12.9$\pm$6.1	 &0.8346	 &5.7	 &45.3$\pm$21.4$\pm$2.8\\
2.1250	 &109	 &309.6$\pm$31.5	 &0.8555	 &9.6	 &70.6$\pm$7.2$\pm$4.9\\
2.1500	 &2.84	 &15.8$\pm$5.9	 &0.8714	 &13.7	 &94.7$\pm$35.4$\pm$7.9\\
2.1750	 &10.6	 &84.5$\pm$15.6	 &0.8835	 &18.8	 &97.3$\pm$18.0$\pm$6.1\\
2.2000	 &13.7	 &137.7$\pm$18.7	 &0.8898	 &21.7	 &105.8$\pm$14.4$\pm$7.8\\
2.2324	 &11.9	 &260.0$\pm$22.3	 &0.8543	 &27.2	 &191.8$\pm$16.5$\pm$14.4\\
2.3094	 &21.1	 &377.0$\pm$26.0	 &0.9465	 &32.6	 &117.8$\pm$8.1$\pm$7.1\\
2.3864	 &22.5	 &573.4$\pm$31.6	 &0.9598	 &37.4	 &144.0$\pm$7.9$\pm$13.2\\
2.3960	 &66.9	 &1841.6$\pm$56.2	 &0.9618	 &38.2	 &152.4$\pm$4.6$\pm$11.7\\
2.5000	 &1.10 	&25.5$\pm$6.9	 	&0.9846	 &43.4	 &110.5$\pm$29.9$\pm$10.1\\
2.6444	 &33.7	 &883.1$\pm$37.5	 &1.0211	 &46.4	 &112.3$\pm$4.8$\pm$7.0\\
2.6464	 &34.0	 &901.3$\pm$37.7	 &1.0217	 &46.5	 &113.4$\pm$4.7$\pm$6.5\\
2.7000	 &1.03	 &26.0$\pm$6.1	 &1.0376	 &48.8	 &100.9$\pm$23.7$\pm$9.4\\
2.8000	 &1.01 	&13.2$\pm$4.5	 	&1.0702	 &47.9	 &51.9$\pm$17.7$\pm$4.7\\
2.9000	 &105	 &2010.8$\pm$54.4	 &1.1013	 &49.2	 &71.7$\pm$1.9$\pm$3.9\\
2.9500	 &15.9	 &282.2$\pm$20.4	 &1.1099	 &48.6	 &66.7$\pm$4.8$\pm$3.7\\
2.9810	 &16.1	 &245.9$\pm$20.0	 &1.1098	 &49.5	 &56.6$\pm$4.6$\pm$3.1\\
3.0000	 &15.9	 &242.6$\pm$18.8	 &1.1064	 &50.0	 &56.1$\pm$4.3$\pm$3.4\\
3.0200	 &17.3	 &253.7$\pm$19.9	 &1.0996	 &50.2	 &54.0$\pm$4.2$\pm$3.1\\
3.0800	 &126	 &1690.8$\pm$50.1	 &1.0065	 &49.7	 &54.4$\pm$1.6$\pm$2.8\\
\hline
 \hline
\end{tabular}

\end{lrbox}
\scalebox{0.75}{\usebox{\tablebox}}

\label{XSectionSummaryPhiKK}
\end{table}

\begin{table*}[htbp]
  \centering
  \caption{Relative systematic uncertainties (in $\%$) for the cross section of $e^{+}e^{-} \to K^{+}K^{-}K^{+}K^{-}$. The uncertainties are associated with the
    luminosity ($\mathcal{L}$), tracking efficiency (Tracking), PID efficiency (PID), fit range (Range), signal and background shape (Sig. shape and Bck. shape),
    the initial-state radiation factor (ISR), the vacuum-polarization correction factor (VP), the weighted detection efficiency ($\mathrm{\epsilon}$),
    MC statistics (MC) and others. The total uncertainty is obtained by summing the individual contributions in quadrature.}

   \begin{lrbox}{\tablebox}
  \begin{tabular}{ccccccccccccc}
  \hline
  \hline
  $\sqrt{s}$ (GeV)  &$\mathcal{L}$    &Tracking   &PID  &Range     &Sig. shape     &Bck. shape &ISR  &VP  &$\mathrm{\epsilon}$    &MC  &Others  & Total\\
    \hline
2.1000	 & 1.0	 & 3.0	 & 3.0	 & 3.2	 & 0.3	 & 3.2	 & 0.1	 & 0.5	 & 2.3	 & 1.2	 & 1.0	 & 6.9\\
2.1250	 & 1.0	 & 3.0	 & 3.0	 & 0.8	 & 1.9	 & 0.1	 & 0.1	 & 0.5	 & 0.6	 & 0.9	 & 1.0	 & 5.1\\
2.1500	 & 1.0	 & 3.0	 & 3.0	 & 3.8	 & 1.6	 & 4.4	 & 0.7	 & 0.5	 & 2.6	 & 0.7	 & 1.0	 & 8.0\\
2.1750	 & 1.0	 & 3.0	 & 3.0	 & 1.9	 & 7.3	 & 0.3	 & 0.3	 & 0.5	 & 1.3	 & 0.6	 & 1.0	 & 8.9\\
2.2000	 & 1.0	 & 3.0	 & 3.0	 & 0.1	 & 0.6	 & 7.6	 & 0.5	 & 0.5	 & 1.1	 & 0.5	 & 1.0	 & 9.0\\
2.2324	 & 1.0	 & 3.0	 & 3.0	 & 0.7	 & 0.1	 & 0.6	 & 0.4	 & 0.5	 & 0.7	 & 0.5	 & 1.0	 & 4.7\\
2.3094	 & 1.0	 & 3.0	 & 3.0	 & 1.4	 & 2.1	 & 4.5	 & 0.4	 & 0.5	 & 0.6	 & 0.4	 & 1.0	 & 6.9\\
2.3864	 & 1.0	 & 3.0	 & 3.0	 & 0.2	 & 0.0	 & 1.2	 & 0.0	 & 0.5	 & 0.5	 & 0.3	 & 1.0	 & 4.7\\
2.3960	 & 1.0	 & 3.0	 & 3.0	 & 3.5	 & 3.5	 & 3.8	 & 0.4	 & 0.5	 & 0.3	 & 0.3	 & 1.0	 & 7.7\\
2.5000	 & 1.0	 & 3.0	 & 3.0	 & 0.8	 & 0.3	 & 2.7	 & 0.3	 & 0.5	 & 2.0	 & 0.3	 & 1.0	 & 5.7\\
2.6444	 & 1.0	 & 3.0	 & 3.0	 & 0.3	 & 0.1	 & 0.7	 & 0.1	 & 0.5	 & 0.3	 & 0.3	 & 1.0	 & 4.6\\
2.6464	 & 1.0	 & 3.0	 & 3.0	 & 0.0	 & 0.1	 & 0.2	 & 0.5	 & 0.5	 & 0.3	 & 0.3	 & 1.0	 & 4.5\\
2.7000	 & 1.0	 & 3.0	 & 3.0	 & 0.2	 & 0.2	 & 7.5	 & 0.3	 & 0.5	 & 1.7	 & 0.3	 & 1.0	 & 8.9\\
2.8000	 & 1.0	 & 3.0	 & 3.0	 & 1.1	 & 1.9	 & 3.8	 & 0.3	 & 0.5	 & 1.8	 & 0.2	 & 1.0	 & 6.5\\
2.9000	 & 1.0	 & 3.0	 & 3.0	 & 0.4	 & 0.2	 & 0.5	 & 0.0	 & 0.5	 & 0.1	 & 0.2	 & 1.0	 & 4.6\\
2.9500	 & 1.0	 & 3.0	 & 3.0	 & 0.9	 & 0.4	 & 0.8	 & 0.3	 & 0.5	 & 0.4	 & 0.3	 & 1.0	 & 4.7\\
2.9810	 & 1.0	 & 3.0	 & 3.0	 & 0.2	 & 0.7	 & 1.6	 & 0.1	 & 0.5	 & 0.4	 & 0.2	 & 1.0	 & 4.9\\
3.0000	 & 1.0	 & 3.0	 & 3.0	 & 0.4	 & 0.7	 & 0.4	 & 0.2	 & 0.5	 & 0.3	 & 0.2	 & 1.0	 & 4.6\\
3.0200	 & 1.0	 & 3.0	 & 3.0	 & 1.9	 & 0.8	 & 1.1	 & 0.0	 & 0.5	 & 0.3	 & 0.2	 & 1.0	 & 5.1\\
3.0800	 & 1.0	 & 3.0	 & 3.0	 & 0.8	 & 0.3	 & 0.1	 & 0.0	 & 0.5	 & 0.1	 & 0.3	 & 1.0	 & 4.6\\
    \hline
    \end{tabular}
    
    \end{lrbox}
    \scalebox{1.0}{\usebox{\tablebox}}
    
    \label{XTotalSystematicKKKK}
\end{table*}


\begin{table*}[htbp]
  \scriptsize
  \centering
  \caption{Summary of relative systematic uncertainties (in $\%$) related to the cross section measurements of $e^{+}e^{-} \to \phi K^{+}K^{-}$.
    See Table~\ref{XTotalSystematicKKKK} for a description of the various items. $\mathcal{B}$ refers to the uncertainty in the branching fraction $\phi\rightarrow K^+K^-$.}
  
   \begin{lrbox}{\tablebox}
  
  \begin{tabular}{ccccccccccccccc}
  \hline
  \hline
  $\sqrt{s}$ (GeV)  &$\mathcal{L}$    &Tracking   &PID        &Kinematic     &Sig. shape     &Bck. shape    &Range  &ISR   &VP   &$\mathrm{\epsilon}$    &MC     &$\mathcal{B}$     &Others  & Total\\
  \hline
2.1000	 &1.0	 &3.0	 &3.0	 &2.0	 &0.7	 &0.0	 &1.2	 &1.3	 &0.5	 &2.4	 &1.3	 &1.3	 &1.0	 &6.1\\
2.1250	 &1.0	 &3.0	 &3.0	 &2.1	 &0.0	 &2.8	 &3.4	 &0.9	 &0.5	 &1.0	 &1.0	 &1.3	 &1.0	 &7.0\\
2.1500	 &1.0	 &3.0	 &3.0	 &2.5	 &0.7	 &5.9	 &1.2	 &0.7	 &0.5	 &1.5	 &0.8	 &1.3	 &1.0	 &8.3\\
2.1750	 &1.0	 &3.0	 &3.0	 &2.2	 &2.2	 &1.2	 &2.2	 &0.2	 &0.5	 &1.2	 &0.7	 &1.3	 &1.0	 &6.3\\
2.2000	 &1.0	 &3.0	 &3.0	 &2.4	 &3.6	 &2.2	 &2.9	 &0.3	 &0.5	 &1.0	 &0.6	 &1.3	 &1.0	 &7.4\\
2.2324	 &1.0	 &3.0	 &3.0	 &2.4	 &5.2	 &0.4	 &0.0	 &0.5	 &0.5	 &0.9	 &0.5	 &1.3	 &1.0	 &7.5\\
2.3094	 &1.0	 &3.0	 &3.0	 &2.3	 &2.5	 &0.8	 &1.0	 &0.1	 &0.5	 &0.8	 &0.5	 &1.3	 &1.0	 &6.0\\
2.3864	 &1.0	 &3.0	 &3.0	 &2.0	 &7.3	 &0.9	 &2.1	 &0.1	 &0.5	 &0.6	 &0.4	 &1.3	 &1.0	 &9.2\\
2.3960	 &1.0	 &3.0	 &3.0	 &1.7	 &5.6	 &0.1	 &1.6	 &0.0	 &0.5	 &0.3	 &0.4	 &1.3	 &1.0	 &7.7\\
2.5000	 &1.0	 &3.0	 &3.0	 &1.7	 &6.7	 &0.0	 &3.3	 &0.1	 &0.5	 &1.1	 &0.4	 &1.3	 &1.0	 &9.1\\
2.6444	 &1.0	 &3.0	 &3.0	 &1.6	 &2.5	 &1.9	 &2.1	 &0.3	 &0.5	 &0.3	 &0.3	 &1.3	 &1.0	 &6.2\\
2.6464	 &1.0	 &3.0	 &3.0	 &1.6	 &2.5	 &0.5	 &0.9	 &0.7	 &0.5	 &0.3	 &0.3	 &1.3	 &1.0	 &5.7\\
2.7000	 &1.0	 &3.0	 &3.0	 &1.6	 &7.7	 &0.0	 &0.0	 &0.0	 &0.5	 &1.3	 &0.3	 &1.3	 &1.0	 &9.3\\
2.8000	 &1.0	 &3.0	 &3.0	 &1.5	 &7.1	 &0.0	 &0.0	 &0.0	 &0.5	 &2.6	 &0.3	 &1.3	 &1.0	 &9.0\\
2.9000	 &1.0	 &3.0	 &3.0	 &1.5	 &2.2	 &0.1	 &0.1	 &0.4	 &0.5	 &0.3	 &0.3	 &1.3	 &1.0	 &5.4\\
2.9500	 &1.0	 &3.0	 &3.0	 &1.3	 &2.2	 &0.3	 &0.7	 &0.1	 &0.5	 &0.8	 &0.3	 &1.3	 &1.0	 &5.5\\
2.9810	 &1.0	 &3.0	 &3.0	 &1.4	 &1.7	 &0.4	 &1.3	 &0.2	 &0.5	 &1.0	 &0.3	 &1.3	 &1.0	 &5.5\\
3.0000	 &1.0	 &3.0	 &3.0	 &1.4	 &0.4	 &1.3	 &3.0	 &0.1	 &0.5	 &0.9	 &0.3	 &1.3	 &1.0	 &6.0\\
3.0200	 &1.0	 &3.0	 &3.0	 &1.4	 &2.7	 &0.8	 &0.8	 &0.2	 &0.5	 &0.8	 &0.3	 &1.3	 &1.0	 &5.8\\
3.0800	 &1.0	 &3.0	 &3.0	 &1.3	 &0.7	 &0.0	 &1.2	 &0.3	 &0.5	 &0.5	 &0.3	 &1.3	 &1.0	 &5.1\\
\hline
  \hline
    \end{tabular}
    
    \end{lrbox}
    \scalebox{1.25}{\usebox{\tablebox}}
    
    \label{XTotalSystematicPhiKK}
\end{table*}

\section{SYSTEMATIC UNCERTAINTY}

Several sources of systematic uncertainties are considered in the
measurement of the Born cross sections. These include the luminosity
measurements, the differences between the data and the MC simulation
for the tracking efficiency, PID efficiency, kinematic fit, the fit
procedure, the MC simulation of the ISR-correction factor and the
vacuum-polarization factor, as well as uncertainties in the branching
fractions of the decays of intermediate states.

(a) \emph{Luminosity:} The integrated luminosity of the data samples
used in this analysis are measured using large-angle Bhabha scattering events,
and the corresponding uncertainties are estimated to be
$1.0\%$~\cite{LuminosityFinal}.

(b) \emph{Tracking efficiency:} The uncertainty of the tracking
efficiency is investigated using a control sample of the $e^{+}e^{-}
\to K^{+} K^{-}\pi^{+}\pi^{-}$ process~\cite{Liudong}. The difference
in tracking efficiency between data and the MC simulation is estimated
to be 1$\%$ per track. Hence, $3.0\%$ is taken as the systematic
uncertainty for the three selected
kaons.

(c) \emph{PID efficiency:} To estimate the uncertainty in the PID efficiency, we study $K^{\pm}$ PID efficiencies with the same control samples as those used in the tracking efficiency. The average difference in PID efficiency between data and the MC simulation is found to be 1$\%$ per charged track. Therefore, $3.0\%$ is taken as the systematic uncertainty for the three selected kaons.

(d) \emph{Kinematic fit:} The uncertainty associated with the kinematic fits comes from the inconsistency of the track helix parameters between data and the MC simulation. The helix parameters for the charged tracks of MC samples are corrected to eliminate the inconsistency, as described in Ref.~\cite{Aixiaocong}, and the agreement of $\chi^{2}$ distributions between data and the MC simulation is significantly improved. We take the differences of the selection efficiencies with and without the correction as the systematic uncertainties.

(e) \emph{Fit procedure:} A fit to mass spectrum of the recoiling
kaon is performed to determine the signal yields of the $e^{+}e^{-} \to
K^{+}K^{-}K^{+}K^{-}$ process, and the two kaon invariant mass
$M(K^{+}K^{-})$ is fitted to determine the number of $e^{+}e^{-} \to
\phi K^{+}K^{-}$ events. The following three aspects are considered
when evaluating the systematic uncertainty associated with the fit
procedure.

(1) {\it Fit range:} The $M(K^{\pm})$ spectrum of the recoiling
kaon is fitted by varying the range from (0.3, 0.7) GeV/$c^2$ to
(0.31, 0.69) GeV/$c^2$.  The $M(K^{+}K^{-})$ spectrum is fitted in the
region from 0.98 to 1.15 GeV/$c^2$. An alternative fit range, from
0.98 to 1.20 GeV/$c^2$, is considered.  The differences between the
yields are treated as the systematic uncertainty from the fit range.

(2) {\it Signal shape:} The signal shape of the mass spectrum of the recoiling kaon is
described by a shape obtained from a MC simulation convolved with a
Gaussian function.  The uncertainty related to this line shape is
estimated with an alternative fit using the same line-shape function,
but fixing the width of the Gaussian function to a value differing by
one standard deviation from the width obtained in the nominal fit. The
signal shape of the $\phi$ is described by a $P$-wave Breit-Wigner
function convolved with a Gaussian function. An alternative fit with
a MC shape convolved with a Gaussian function is performed. The
difference in yield between the various fits is considered as the
systematic uncertainty from the signal shape.

(3) {\it Background shape:} The background shape of the mass spectrum for the recoiling kaon is described as a second-order Chebyshev polynomial function.
A fit with a first-order Chebyshev polynomial function for the background shape is used to estimate its uncertainty. The background shape for $\phi$-mass distribution
is described by an ARGUS function. The fit with a function of $f(M) = (M-M_{a})^{c}(M_{b}-M)^{d}$, where, $M_{a}$ and $M_{b}$ are the lower and upper edges of the mass distribution, is used to estimate this uncertainty.

(f) \emph{ISR factor:} The cross section is measured by iterating until $(1+\delta^{\mathrm{r}})\epsilon$ converges, and the difference between the last two iterations is taken as the systematic uncertainty associated with the ISR-correction factor.

(g) \emph{VP factor:} The uncertainty on the calculation of the VP factor is 0.5$\%$~\cite{VP}.

(h) \emph{Branching fraction:} The experimental uncertainties in the branching fraction for the process $\phi \to K^{+}K^{-}$ are taken from the PDG~\cite{PDG}.

(i) \emph{Weighted detection efficiency:} The detection efficiencies obtained in different processes are combined using the previously-described method. The combined uncertainty is calculated by accounting for the statistical variation, by one standard deviation, of the signal yields.

To obtain a reliable detection efficiency of $e^{+}e^{-} \to \phi K^{+}K^{-}$, the PHSP MC sample is weighted to match the distribution of the background-subtracted data. To consider the effect on the statistical fluctuations of the signal yield in the data, a set of toy-MC samples, which are produced by sampling the signal yield and its statistical uncertainty of the data in each bin, are used to estimate the detection efficiencies. 

(j) \emph{MC statistics:} The uncertainty is estimated by the number of the generated events, whereby the weighting factor has been taken into account.

(k) \emph{Other systematic uncertainties:} Other sources of systematic uncertainties include the trigger efficiency, the determination of the start time of an event,
and the modeling of the final-state radiation in the simulation. The total systematic uncertainty due to these sources is estimated to be less than $1.0\%$.
To be conservative, we take 1.0$\%$ as its systematic uncertainty.

Assuming all of the above systematic uncertainties, shown in Tables~\ref{XTotalSystematicKKKK} and~\ref{XTotalSystematicPhiKK}, are independent,
the total systematic uncertainties are obtained by adding the individual uncertainties in quadrature.

\section{Summary and Discussion}

In summary, using data collected with the BESIII detector taken at
twenty c.m.\ energies from 2.100 to 3.080~GeV, we present measurements
of the processes $e^{+}e^{-} \to K^{+}K^{-}K^{+}K^{-}$ and $\phi
K^{+}K^{-}$ and we obtain the corresponding Born cross sections. The
Born cross sections of the process $e^{+}e^{-} \to
K^{+}K^{-}K^{+}K^{-}$ are in good agreement with the results by BABAR,
but with improved precision. The Born cross sections for the channel $e^{+}e^{-} \to \phi K^{+}K^{-}$ are measured for the first time at twenty energy points. Both data sets reveal anomalously high cross sections at $\sqrt{s} = 2.232$~GeV.

A previous analysis on a much smaller dataset~\cite{JpsiToPhiKK} has demonstrated that the $K^{+}K^{-}K^{+}K^{-}$ final state exhibits resonant substructure. It is difficult to disentangle these contributions from other final states, and we make no attempt to do so.

By examining the $\phi K^{+}K^{-}$ cross section as a function of
c.m.\ energy, an enhancement at $\sqrt{s}$ = 2.232~GeV, \emph{i.e.} near the
$\Lambda \overline{\Lambda}$ production threshold, is observed. The
cross section of $e^{+}e^{-} \to \Lambda \overline{\Lambda}$ is also
found to be anomalously high at the
threshold~\cite{LambdaAntiLambdaXiaorongZhou}. In the case of charged
baryons one would expect a Coulomb enhancement factor, which, however,
is absent in the of the electrically-neutral $\Lambda$. It has been
suggested that a narrow resonance, very close to the threshold, might
provide an explanation~\cite{FormFactorOfLambda}.  BABAR has observed
an enhancement at 2.175~GeV and a sharp peak at 2.3~GeV,
corresponding to $\phi K^{+} K^{-}$ final states with $K^{+}K^{-}$
invariant masses smaller than 1.06~GeV/$c^2$ and within a mass
interval of 1.06$-$1.2~GeV/$c^{2}$, respectively.  The intriguing
$\phi(2170)$ resonance~\cite{X2170} has a relatively wide width and it
is very close to the kinematical threshold, but not close enough to be
related to the observed anomaly. Alternatively, the enhancement at
2.232~GeV could be explained by an interference effect of different
resonances.  More data in the vicinity would be helpful to understand
the anomaly.

\section{acknowledgments}

The BESIII collaboration thanks the staff of BEPCII and the IHEP computing center and the supercomputing center of USTC for their strong support. This work is supported in part by National Key Basic Research Program of China under Contract No. 2015CB856700; National Natural Science Foundation of China (NSFC) under Contracts Nos. 11625523, 11635010, 11735014, 11425524, 11335008, 11375170, 11475164, 11475169, 11605196, 11605198, 11705192; National Natural Science Foundation of China (NSFC) under Contract No. 11835012; the Chinese Academy of Sciences (CAS) Large-Scale Scientific Facility Program; Joint Large-Scale Scientific Facility Funds of the NSFC and CAS under Contracts Nos. U1532257, U1532258, U1732263, U1832207, U1532102, U1732263, U1832103; CAS Key Research Program of Frontier Sciences under Contracts Nos. QYZDJ-SSW-SLH003, QYZDJ-SSW-SLH040; 100 Talents Program of CAS; INPAC and Shanghai Key Laboratory for Particle Physics and Cosmology; German Research Foundation DFG under Contract No. Collaborative Research Center CRC 1044, FOR 2359; Istituto Nazionale di Fisica Nucleare, Italy; Koninklijke Nederlandse Akademie van Wetenschappen (KNAW) under Contract No. 530-4CDP03; Ministry of Development of Turkey under Contract No. DPT2006K-120470; National Science and Technology fund; The Knut and Alice Wallenberg Foundation (Sweden) under Contract No. 2016.0157; The Swedish Research Council; U. S. Department of Energy under Contracts Nos. DE-FG02-05ER41374, DE-SC-0010118, DE-SC-0012069, DE-SC-0010504; University of Groningen (RuG) and the Helmholtzzentrum fuer Schwerionenforschung GmbH (GSI), Darmstadt



\end{document}